# Radiation-induced Acoustic Signal Denoising using a Supervised Deep Learning Framework for Imaging and Therapy Monitoring

**Zhuoran Jiang** [1, 2, +], **Siqi Wang** [3, +], **Yifei Xu** [3], **Leshan Sun** [3], **Gilberto Gonzalez** [4], **Yong Chen** [4], **Q. Jackie Wu** [1, 2], **Liangzhong Xiang** [3, 5, 6, *], **Lei Ren** [7, *]

[1] Medical Physics Graduate Program, Duke University, Durham, NC, 27705, USA

[2] Department of Radiation Oncology, Duke University Medical Center, Durham, NC, 27710, USA

[3] Department of Biomedical Engineering, University of California, Irvine, California 92617, USA

[4] Department of Radiation Oncology, University of Oklahoma Health Sciences Center, Oklahoma City, OK, 73104, USA

[5] Department of Radiological Sciences, University of California, Irvine, CA 92697, USA

[6] Beckman Laser Institute & Medical Clinic, University of California, Irvine, Irvine, CA 92612, USA

[7] Department of Radiation Oncology, University of Maryland, Baltimore, MD, 21201, USA

[+] Contributed equally

[*] Corresponding Authors

**Emails:** lren@som.umaryland.edu (Lei Ren) and liangzhx@hs.uci.edu (Liangzhong Xiang)

**Abstract**

Radiation-induced acoustic (RA) imaging is a promising technique for visualizing the invisible radiation energy deposition in tissues, enabling new imaging modalities and real-time therapy monitoring. However, RA imaging signal often suffers from poor signal-to-noise ratios (SNRs), thus requiring measuring hundreds or even thousands of frames for averaging to achieve satisfactory quality. This repetitive measurement increases ionizing radiation dose and degrades the temporal resolution of RA imaging, limiting its clinical utility. In this study, we developed a general deep inception convolutional neural network (GDI-CNN) to denoise RA signals to substantially reduce the number of frames needed for averaging. The network employs convolutions with multiple dilations in each inception block, allowing it to encode and decode signal features with varying temporal characteristics. This design generalizes GDI-CNN to denoise acoustic signals resulting from different radiation sources. The performance of the proposed method was evaluated using experimental data of X-ray-induced acoustic, protoacoustic, and electroacoustic signals both qualitatively and quantitatively. Results demonstrated the effectiveness of GDI-CNN: it achieved X-ray-induced acoustic image quality comparable to 750-frame-averaged results using only 10-frame-averaged measurements, reducing the imaging dose of X-ray-acoustic computed tomography (XACT) by 98.7%; it realized proton range accuracy parallel to 1500-frame-averaged results using only 20-frame-averaged measurements, improving the range verification frequency in proton therapy from 0.5Hz to 37.5Hz; it reached electroacoustic image quality comparable to 750-frame-averaged results using only a single frame signal, increasing the electric field monitoring frequency from 1 fps to 1k fps. Compared to lowpass filter-based denoising, the proposed method demonstrated considerably lower mean-squared-errors, higher peak-SNR, and higher structural similarities with respect to the corresponding high-frame-averaged measurements. The proposed deep learning-based denoising framework is a generalized method for few-frame-averaged acoustic signal denoising, which significantly improves the RA imaging's clinical utilities for low-dose imaging and real-time therapy monitoring.

# 1. Introduction

Radiation deposits energy when it travels through human bodies and interacts with tissue atoms through various mechanisms, which enables numerous medical applications. Radiation has been widely used for diagnostic imaging. For example, low-energy (kV) X-ray has been widely used in mammography or computed tomography (CT) for tissue characterization and disease diagnosis. Radiation has also been widely used for cancer treatment. For example, ionizing radiation has been extensively developed in radiation therapy using photon, electron, or proton beams to kill cancer cells. Such radiation particles damage the DNA structures in living cells directly or indirectly with free radicals and ions, preventing them from dividing and growing. Non-ionizing radiation has also been developed for cancer treatments, such as radiofrequency ablation and electroporation.

Despite the success of employing radiation in medical applications, the efficacy of using radiation for imaging and treatment remains to be improved. One critical issue is the lack of visibility of radiation energy deposition in tissues. For imaging, x-ray-based imaging modalities have historically relied on detecting penetrating radiation to visualize the internal tissues. This mechanism renders the absorbed dose useless, and necessitates higher dose for imaging and full-view acquisition for 3-dimensional (3D) reconstruction. For treatment, the efficacy of radiation depends on the energy deposition accuracy, which, however, can be affected by many factors such as patient positioning, anatomy changes, and dose calculation uncertainties. Currently, there is a lack of effective tools to verify the internal radiation dose deposition in real-time during the treatment to ensure that the target receives the prescribed dose and the healthy tissues are spared.

Various techniques are being investigated to address the challenges above for radiation-based imaging and treatment. One promising direction is the radiation-induced acoustic (RA) imaging [1,2], which involves detecting acoustic signals generated from pulsed radiation beams and can be detected by ultrasound transducers. The pressure distribution reconstructed from these acoustic signals linearly correlates to the radiation energy deposition, making RA imaging a valuable tool for diagnostic imaging and therapy monitoring. For example, X-ray-induced acoustic computed tomography (XACT) is a novel imaging technique that excites the tissues with pulsed kV X-ray beams to induce acoustic signals[3-9]. It leverages the high contrast of tissue's radiation absorption properties to reveal fine structures deep within the body. Compared to X-ray imaging and CT, XACT features much lower imaging dose and can reconstruct 3D images from a single-view acquisition. Similar to XACT, proton pencil beams in proton therapy can also induce detectable acoustic signals, giving rise to the protoacoustic imaging technique for real-time proton dose verification[10-17]. In addition, studies have found that electric fields used in the electroporation for tumor ablation can also induce acoustic signals due to the electrical energy deposition [18-22]. This phenomenon led to the development of the electroacoustic tomography (EAT) technique, allowing the electroporation process to be monitored in real time.

Despite the promise of these novel RA techniques, a significant challenge with RA imaging is its low signal-to-noise ratio (SNR) in the acoustic signal. The detected acoustic signals are usually contaminated with background noises, such as electronic and system thermal noises, resulting in poor SNR and limiting the techniques' utility. Averaging is the most common way to improve the signal SNR by eliminating uncorrelated random noise[23]. Typically, hundreds to thousands of frames are required to be averaged to achieve satisfactory SNR, which significantly prolongs the imaging time and degrades the temporal resolution for therapy monitoring and moving-target imaging. In the case of ionizing radiation-based techniques such as XACT, the use of averaging requires the acquisition of a large number of frames, leading to a considerable increase in the imaging dose. Filtering is another widely used technique to denoise acoustic signals. It involves decomposing signals in different domains (e.g., frequencies and wavelet coefficients) and ruling out noises with thresholds [24-28]. The main limitation of this method lies in the tradeoff between reducing the residual noises and preserving the true acoustic signals.

In the recent decade, deep learning has demonstrated superior performance in various medical imaging tasks [29-35]. Its applications have also been explored in acoustic imaging. Antholzer et al. [36] developed a deep convolutional neural network (CNN) to reconstruct accurate photoacoustic images from sparsely sampled data. Hariri et al. [37] used a multi-level wavelet CNN to improve the low-fluence photoacoustic image quality. Jiang et al. [38] developed a cascaded CNN to correct the limited-view artifacts in the matrix array-based proton acoustic imaging. In these studies, deep learning models were used as a post-processing technique, which reduced the artifacts and enhanced the quality of the images reconstructed by conventional algorithms. Earlier research [35,39,40] has indicated that the models can achieve better performance by utilizing advanced signal preprocessing methods to improve the input quality. Several studies have explored this preprocessing area for acoustic imaging. Gutta et al. [41] used a fully-connected network to map the limited bandwidth to the full bandwidth photoacoustic signals. Awasthi et al. [42] employed a CNN to remove the gaussian noise in photoacoustic signals in a simulation study. And for the few-average signal denoising, Wang et al. [43] proposed a dictionary learning-based workflow for both signal denoising and image enhancement in the photoacoustic microscopy (PAM), which reduced the laser fluence by 5 times without compromising the image quality. However, the proposed method used the K-SVD algorithm for sparse coding, assuming that the structures in the underlying signals have sparse representations. This assumption may not be true for images with complex structures and thus limits its applications.

In this study, we developed a general deep-learning model to generate accurate and high-SNR acoustic signals from few-frame-averaged measurements. Compared to the previous dictionary learning-based method [43], the proposed deep learning-based method does not assume the properties of underlying signals and images, making it more generalizable for different tasks. To handle signals with diverse time windows, convolutions with multiple dilations are used in each inception block in the proposed model. This design allows for the encoding and decoding of signal features at different temporal scales. The model's performance was evaluated using the experimental data of kV X-ray-induced acoustic, protoacoustic, and electroacoustic signals, which have potential applications in (1) low-dose imaging with XACT, (2) real-time range verification in proton therapy using protoacoustic imaging, and (3) real-time therapy monitoring for electroporation using EAT. Results showed that the proposed model realized high-SNR signals from extremely few-frame acquisitions for all modalities, demonstrating the model's excellent generalizing ability across various RA techniques using different radiation sources.

This major advance can greatly improve the clinical utilities of these novel radiation-induced acoustic techniques for low-dose diagnostic imaging and real-time treatment verification.

## 2. Methods

### 2.1 Problem Formulation

Let $x \in R^{C \times T}$ be the real-valued RF signals acquired with a transducer having $C$ channels in $T$ timesteps, and $y \in R^{C \times T}$ be the corresponding ground truth signals. Then the problem can be formulated as finding a denoising pattern between $x$ and $y$ so that

$$arg\min_{f} \| f(x) - y \|_2^2$$

where $f$ denotes the denoising function estimated by a deep learning model in this study.

## 2.2 Deep Learning-based RF Acoustic Signal Denoising

In this study, a general deep inception CNN (GDI-CNN) was used for the RF acoustic signal denoising. The few-averaged noisy signals are used as the input to the model, which generates the high-SNR denoised signals.

The proposed GDI-CNN utilizes the encoder-decoder architecture. In the encoder branch, seven inception blocks are stacked to extract high-dimensional hidden features from the noisy input signals. In the decoder branch, seven inception blocks are stacked to decode the features to generate the final denoised signals. In each inception block, seven convolutional layers with increasing dilation rates are used to capture features at different temporal resolutions. Compared to the original inception network [44], this study uses convolutions with different dilations, rather than different filter sizes, to further enlarge the inception fields while significantly reducing the computational assumptions. Besides, residual connections bridge the inception blocks in the decoder with the mirrored ones in the encoder to facilitate the training of deep networks. Fig. 1 shows the structure of the proposed general deep inception convolutional neural network (GDI-CNN). Source codes of the model implementation will be available publicly upon the acceptance of the manuscript.

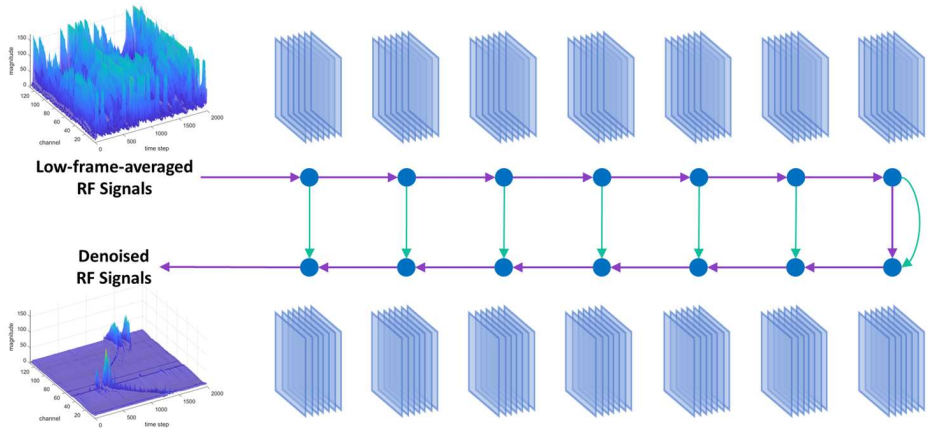

Figure 1. Structure of the proposed general deep inception convolutional neural network (GDI-CNN). It takes few-averaged signal as input, extracts and decodes hidden features with stacked inception convolutions (purple arrows → ) and residual connections (green arrows → ), and generates denoised signal in the end.

## 3. Experiment Design
### 3.1 Data acquisition

In this study, radiation-induced acoustic data of various modalities were collected to verify the proposed method's performance in few-frame-averaged signal denoising. The experiment setup of the X-ray-induced acoustic, protoacoustic, and electroacoustic data collection is illustrated in Fig. 2. Details are described in sections 3.1.1 – 3.1.3.

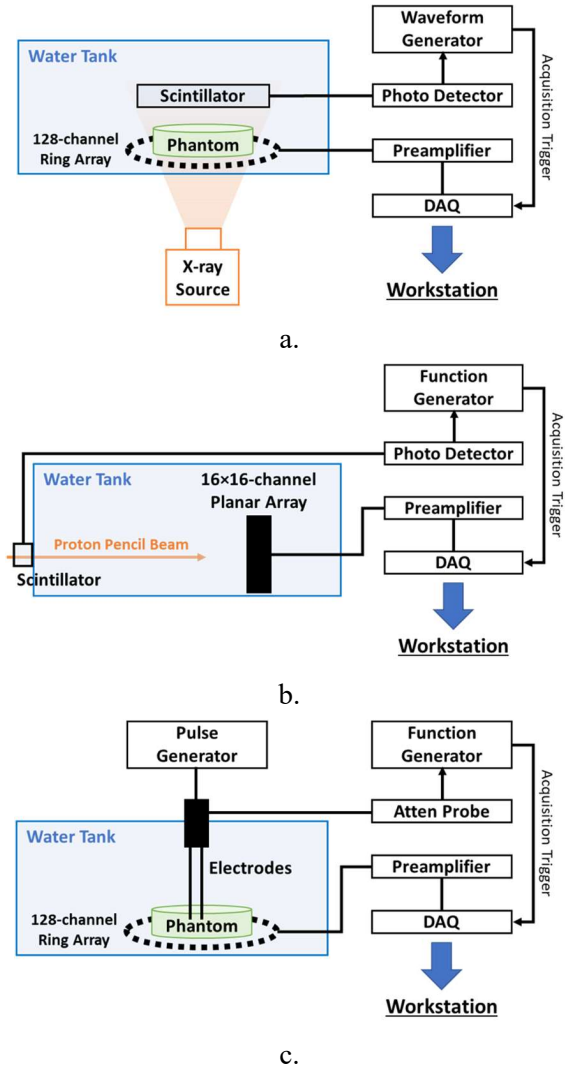

Figure 2. Illustration of the experiment setup of (a) X-ray-induced acoustic data collection, (b) protoacoustic data collection, and (c) electroacoustic data collection.

### 3.1.1 X-ray-induced Acoustic Data

X-ray acoustic signals were generated by irradiating agar phantoms using a portable X-ray source as shown in Fig. 2 (a). Specifically, a battery-powered pulsed X-ray source (XR200, Golden Engineering, IN, USA) operated at 150kVp was used to generate 50-nanosecond X-ray pulses (cone angle: 40°) at a repetition rate of 20 Hz. X-ray beams were projected from the bottom of a water tank, in which a target-holding phantom was placed along the beam trail. The phantoms were made of 3% agar-water solution (BactoTM, Becton, Dickinson and Company, NJ, USA), containing various targets made of 1/16-inch and 1/8-inch-thick lead wires.

The acoustic signals were collected by a ring-shaped 128-channel ultrasound transducer array (PA probe, Doppler Co. Limited, Guangzhou, China; central frequency: 5MHz, bandwidth: ≥ 60%). The detector was placed in the water tank, surrounding the phantom. To trigger the data acquisition, a Ce: Lu2SiO5 crystal (MTI Corporation, CA, USA) was used to scintillate at a wavelength of 418 nm upon incident with stray X-rays. The scintillator was connected to a photodetector (PDA100A, Thorlabs, NJ, USA) for light-to-volt conversion and followed with a waveform generator (33600A, Keysight, CA, USA) to relay the trigger signal to the 128-channel data acquisition system (SonixDAQ, DK Medical, Canada). Raw acoustic signals were conditioned by a low-noise preamplifier (TomoWave Laboratories, Inc., TX,USA). For each trigger, acoustic signals were collected for 67.5 μs (sampling rate: 40 MHz), forming a frame. For each X-ray acoustic data acquisition, 750 frames were collected to ensure the SNR.

In this experiment, 60 X-ray-induced acoustic data sets with different target sizes and locations were collected for model training and testing.

### 3.1.2 Protoacoustic Data

Protoacoustic signals were generated by irradiating a water tank using a clinical synchrocyclotron as shown in Fig. 2 (b). Specifically, a clinical synchrocyclotron Mevion s250i (Mevion Medical Systems, MA, USA) was used to deliver proton beams (energy: 107.6 MeV, proton range: 8.69 cm, beam pulse width: 4μs, uncollimated Bragg Peak) at a repetition rate of 750 Hz. The water tank has a dimension of 32 cm × 32 cm × 40 cm and was placed on the treatment couch.

The acoustic signals were collected by a custom planar 16×16-channel ultrasound transducer array (Matrix probe, Doppler Co. Limited, Guangzhou, China; central frequency: 1MHz, bandwidth: ≥60%, element size: 3×3mm$^2$). The detector was placed in the water tank at a depth of 12.7 cm with respect to the proton beam entrance. To trigger the data acquisition, a scintillation crystal (BC408, Epic-Crystal, Guangzhou, China) was attached to the beam entrance window to detect the proton beam passage. The scintillator output was connected to a function generator (SDG-2122X, SIGLENT, OH, USA) to relay the trigger signal to the data acquisition system (Legion ADC, Photosound Technologies, USA), which was housed in a Borated Polyethylene

(BPE) shielded box for neutron shielding. For each trigger, acoustic signals were collected for 67.5 µs (sampling rate: 40 MHz), forming a frame. For each protoacoustic data acquisition, 1500 frames were collected to ensure the SNR.

In this experiment, 5 protoacoustic data sets with various proton beam coordinates were collected for the model training and testing.

### 3.1.3 Electroacoustic Data

Electroacoustic signals were generated by irradiating agar phantoms using an electrical pulse generator as shown in Fig. 2 (c). Specifically, a custom nanosecond electrical pulse generator (VilniusTECH, Lithuania) was used to deliver 100-nanosecond electrical pulses (repetition rate: 1000 Hz) via several tungsten electrodes (573400, A&M Systems, Carlsborg, USA) to an agar phantom in a water tank. The voltage on the electrodes was set to 100 Volt, resulting in an electric field of hundreds of volts per centimeter. A multi-electrode holder was 3D-printed to enable different electrode arrangements. Distances between electrodes varied between 5 mm to 15 mm.

The acoustic signals were collected using the same 128-channel ring array mentioned in section 3.1.1. The detector was placed in the water tank, surrounding the phantom. A polyethylene film was used to isolate the electrodes from the ultrasound transducer to reduce the direct impact of the high-voltage pulsed electric field on the piezoelectric ultrasound probe surface. To trigger the data acquisition, a parallelly connected high-voltage probe (P4250, Keysight Technologies, USA) was used to detect the pulse generation directly from the output. The other ends of the probes were connected to a function generator to relay the trigger signal to the 128-channel data acquisition system (SonixDAQ, DK Medical, Canada). Raw electroacoustic signals were boosted by a custom low-noise preamplifier. For each trigger, acoustic signals were collected for 67.5 µs (sampling rate: 40 MHz), forming a frame. For each electroacoustic data acquisition, 750 frames were collected to ensure the SNR.

In this experiment, 52 electroacoustic data sets from various electrode arrangements were collected for model training and testing.

### 3.2 Model Training

The proposed model was trained in a modality-specific method. For each imaging technique, a model was trained and tested using the corresponding dataset. Specifically, for the XACT, 60 data were acquired, of which 80%, 10%, and 10% were used for model training, validation, and testing, respectively. For each data, 10 consecutive frames were averaged as the input to the model, and the 750-frame-averaged signals were used as the "ground truth" signals. As a result, 4050 samples were used for model training and 450 samples were used for testing. For the protoacoustic, five data sets were acquired. The leave-one-out strategy was used for the model training and testing due to the limited dataset. For each data, 20 consecutive frames were averaged as the input to the model, and the 1500-frame-averaged signals were used as ground truth. As a result, 300 samples were used for model training and 75 samples were used for testing. For the EAT, 52 data were acquired, of which 43 data were used for model training, 3 for validation, and 6 for testing. For each data, a single frame was used as the model input, and the 750-frame-averaged signals were used as the ground truth. To balance the computing time and model performance, one frame out of every ten was used for denoising. As a result, 3450 samples were used for model training and 450 samples were used for testing.

To improve the learning target ("ground truth" signal) quality, for the X-ray-induced acoustic and electroacoustic signals, envelope was performed on the 750-frame-averaged signals; and for the protoacoustic data, the discrete wavelet transform (DWT) was used to further denoise the 1500-frame-averaged signals, which uses a scaled and predefined wavelet and scaling functions to convolute with signals. For DWT, we implemented the 'Coiflets' wavelet filter (order = 5) [27] and the 'sqtwolog' threshold selection [45] using the MATLAB wavelet toolbox.

During the model training, the few-frame-averaged signal was used as input to the model, whose weights were optimized by minimizing the mean squared errors (MSE) between the denoised signal and the corresponding ground truth signal. The optimizer was set to "Adam" [46] with a learning rate of 0.001. The batch size was set to 1 to account for the memory.

### 3.3 Model Evaluation

The signals denoised by the proposed model were compared to the "ground truth" signals. Meanwhile, the widely used lowpass filtering was also included in this study as a comparison baseline. The cutoff frequency for the lowpass filter was set to 6MHz, 200kHz, and 6MHz for X-ray-induced acoustic, protoacoustic, and electroacoustic signals, respectively. The lowpass filter was implemented using the MATLAB signal processing toolbox.

To further demonstrate the values of the denoising, acoustic images were reconstructed using the back-projection algorithm from the denoised signals, and were then compared to those reconstructed from the ground truth signals. An overall workflow for the model evaluation is shown in Fig. 3.

The results were evaluated both qualitatively and quantitatively using metrics of MSE, peak signal-to-noise ratio (PSNR), and structural similarity index (SSIM).

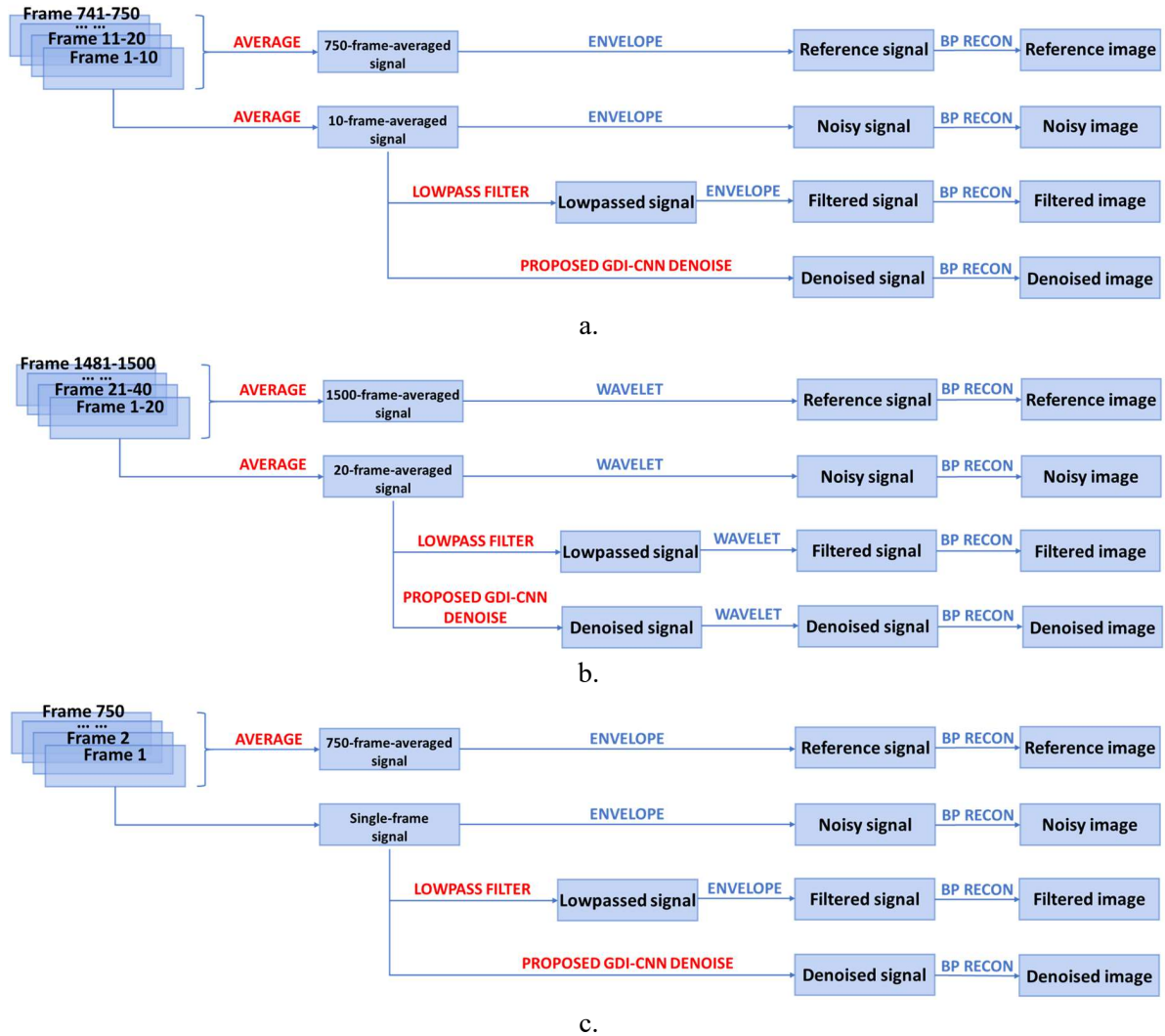

Figure 3. Evaluation workflow of (a) X-ray-induced acoustic data, (b) protoacoustic data, and (c) electroacoustic data. Reference signals/images are used as the ground truth for result evaluation.

## 4. Results

### 4.1 Radiation-induced Acoustic Signal Denoising

Fig. 4 shows a representative 10-frame-averaged X-ray-induced acoustic data. Due to the few-frame average, the RF signals are severely contaminated by background noises. The lowpass filter partially removes the noises but still yields poor SNR. Our proposed method effectively removes the noises while accurately preserving the signals, showing very good agreement with the reference 750-frame-averaged signals. The acoustic image reconstruction further demonstrates the effectiveness of the proposed denoising framework. The targets in the agar phantom are completely diminished in the few-frame-averaged reconstruction, and can be hardly distinguished in the lowpassed reconstruction. In contrast, the image reconstructed from the proposed denoised signals has clear and accurate structures, which has high similarity to the reference image reconstructed from the 750-frame-averaged signals.

Fig. 5 shows a representative 20-frame-averaged protoacoustic data. Signals cannot be distinguished from the few-frame-averaged signals. The lowpass filter partially removes the noises but the signals are still highly contaminated. The proposed method effectively removes the noises while preserving the signals. The acoustic image reconstructed from the few-frame-averaged signals has poor image quality with severe artifact contamination, and the one from lowpass-filtered signals shows obvious artifacts and errors. In contrast, the image reconstructed from the proposed denoised signals shows a clear and accurate proton range, agreeing very well with the 1500-frame-averaged results.

Fig. 6 shows representative single-frame electroacoustic data. Similar to the X-ray-induced acoustic results, the proposed denoising method also considerably improved the SNR of the electroacoustic signals. The reconstructed acoustic images further confirm the effectiveness of the proposed method.

Table 1 and Table 2 show the quantitative results in both acoustic signal and image domains. Compared to the lowpass-filtered few-frame-averaged results, the signals/images denoised by the proposed method showed substantially lower intensity error (indicated by lower MSE), higher SNR (indicated by high PSNR), and higher structural similarities (indicated by higher SSIM) with respect to the corresponding high-frame-averaged reference results for all the X-ray-induced acoustic, protoacoustic, and electroacoustic data. Quantitative results agreed with the qualitative evaluation.

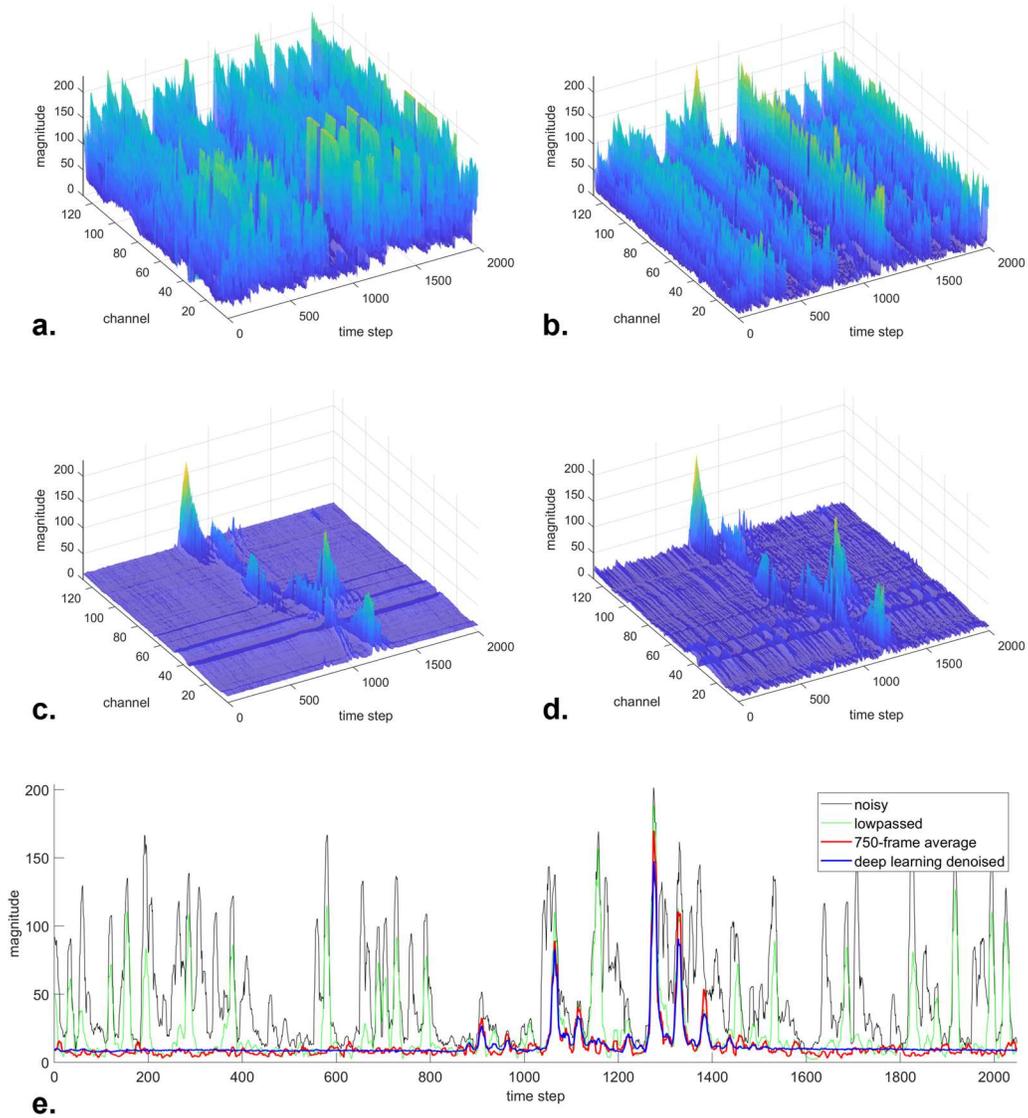

Figure 4A. X-ray-induced acoustic signals of (a) 10-frame-averaged data, and (b) the data denoised by lowpassed filter, (c) the data denoised by the proposed deep learning framework, and (d) the 750-frame-averaged data. (e) shows the signal from an example channel of transducer.

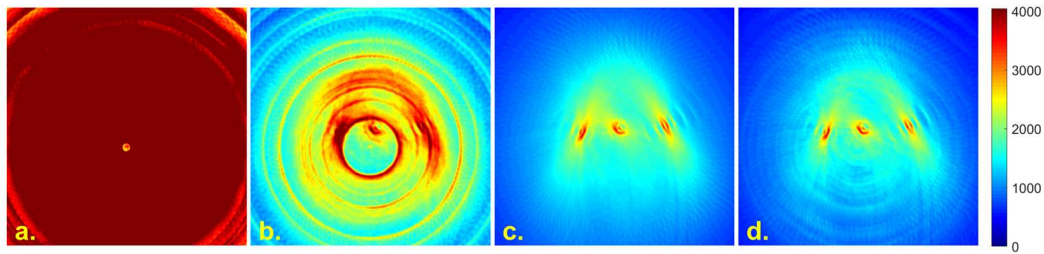

Figure 4B. X-ray-induced acoustic images reconstructed from signals of (a) 10-frame-averaged data, and (b) the data denoised by lowpassed filter, (c) the data denoised by the proposed deep learning framework, and (d) the 750-frame-averaged data.

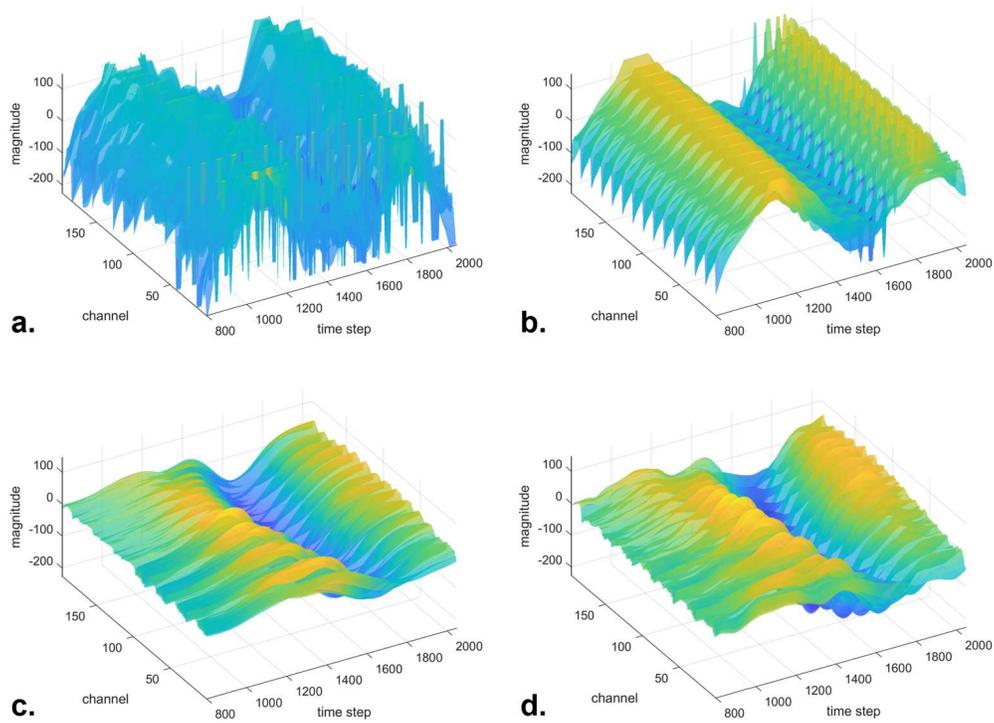

Figure 5A. Protoacoustic signals of (a) 20-frame-averaged data, and (b) the data denoised by lowpassed filter, (c) the data denoised by the proposed deep learning framework, and (d) the 1500-frame-averaged data.

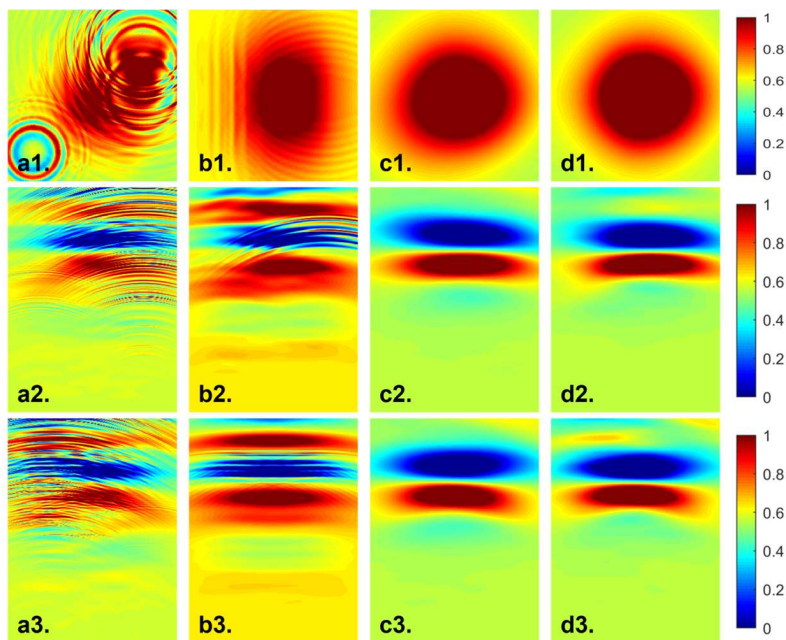

Figure 5B. Protoacoustic images reconstructed from signals of (a) 20-frame-averaged data, and (b) the data denoised by lowpassed filter, (c) the data denoised by the proposed deep learning framework, and (d) the 1500-frame-averaged data, respectively, in axial, coronal, and sagittal views.

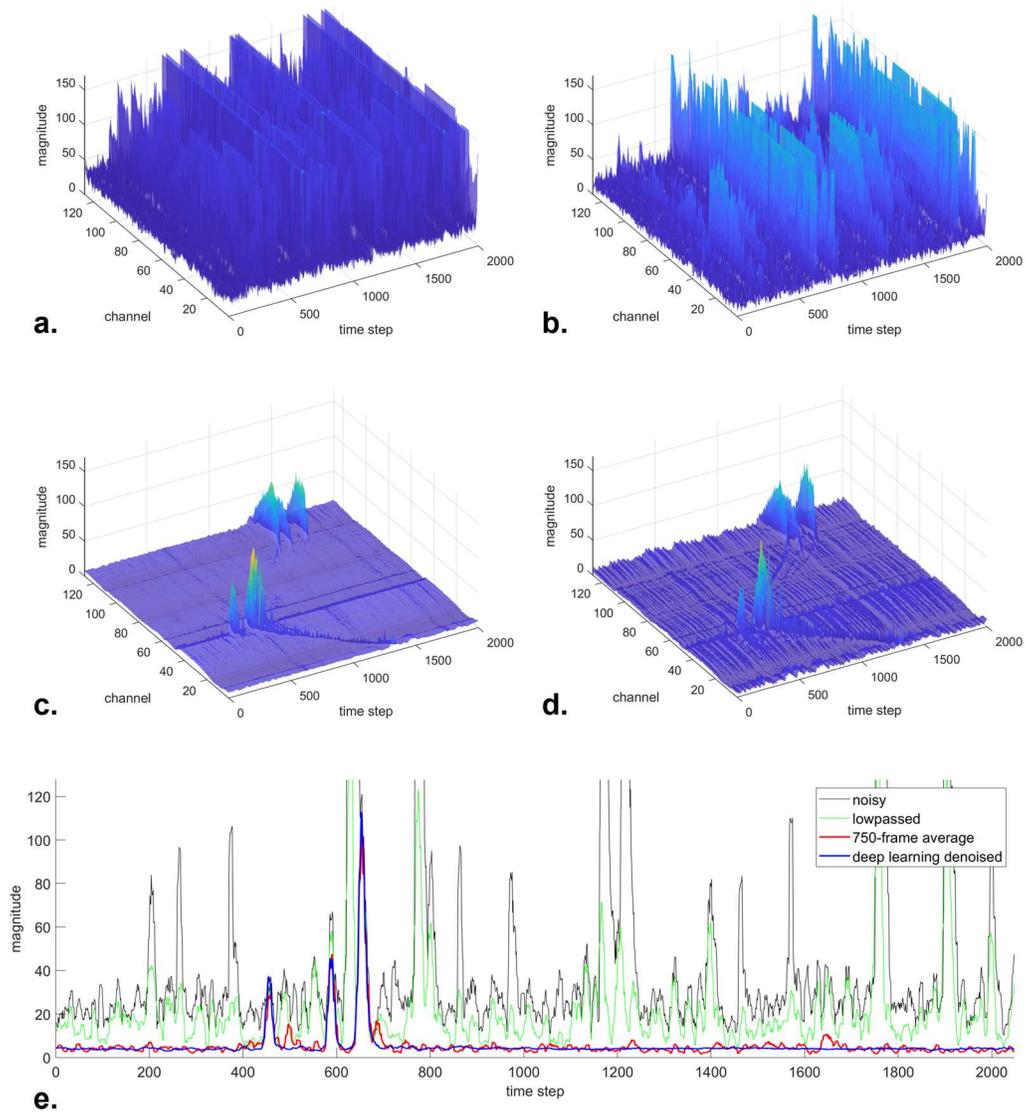

Figure 6A. Electroacoustic signals of (a) single-frame data, and (b) the data denoised by lowpassed filter, (c) the data denoised by the proposed deep learning framework, and (d) the 750-frame-averaged data. (e) shows the signal from an example channel of transducer.

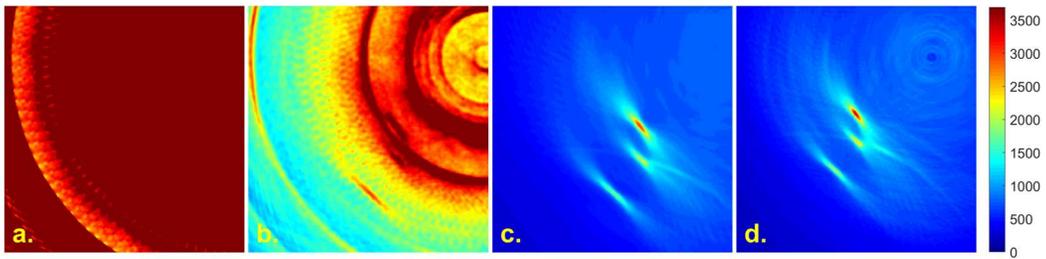

Figure 6B. Electroacoustic images reconstructed from signals of (a) single-frame data, and (b) the data denoised by lowpassed filter, (c) the data denoised by the proposed deep learning framework, and (d) the 750-frame-averaged data.

Table 1. Quantitative analysis of radio-frequency acoustic signals.

| Imaging Techniques | Metrics | Values* | |
| --- | --- | --- | --- |
| | | Filtered | **Proposed** |
| X-ray acoustic | MSE | 0.0099 ± 0.0053 | **0.0026 ± 0.0042** |
| | PSNR | 20.934 ± 3.2003 | **28.972 ± 4.9838** |
| | SSIM | 0.5851 ± 0.0961 | **0.8092 ± 0.1242** |
| Protoacoustic | MSE | 0.0346 ± 0.0202 | **0.0062 ± 0.0013** |
| | PSNR | 15.269 ± 2.3976 | **22.191 ± 0.8483** |
| | SSIM | 0.3690 ± 0.0368 | **0.7774 ± 0.0187** |
| Electroacoustic | MSE | 0.0068 ± 0.0014 | **0.0004 ± 0.0001** |
| | PSNR | 21.754 ± 0.9117 | **34.645 ± 1.9551** |
| | SSIM | 0.5934 ± 0.0491 | **0.8940 ± 0.0165** |

* Values are calculated with data normalized to [0, 1], and expressed as mean ± std.

Table 2. Quantitative analysis of reconstructed images.

| Imaging Techniques | Metrics | Values* | |
| --- | --- | --- | --- |
| | | Filtered | **Proposed** |
| X-ray acoustic | MSE | 0.0114 ± 0.0133 | **0.0081 ± 0.0175** |
| | PSNR | 21.822 ± 4.7896 | **26.955 ± 6.7584** |
| | SSIM | 0.7769 ± 0.1046 | **0.9266 ± 0.0803** |
| Protoacoustic | MSE | 0.0178 ± 0.0148 | **0.0016 ± 0.0002** |
| | PSNR | 18.655 ± 3.0915 | **27.857 ± 0.6377** |
| | SSIM | 0.8313 ± 0.0602 | **0.9738 ± 0.0028** |
| Electroacoustic | MSE | 0.0321 ± 0.0218 | **0.0002 ± 0.0002** |
| | PSNR | 16.385 ± 4.0832 | **38.432 ± 5.2358** |
| | SSIM | 0.6307 ± 0.1661 | **0.9837 ± 0.0081** |

* Values are calculated with data normalized to [0, 1], and expressed as mean ± std.

## 4.2 Runtime

The proposed deep learning framework was implemented using the Keras framework with the TensorFlow backend. The model training and testing were performed on a computer equipped with a CPU of Intel Xeon and 32GB memory and a GPU of NVIDIA Titan RTX (24GB memory). The denoising process is fully automated, which takes about 0.25 seconds for 128-channel X-ray-induced acoustic and electroacoustic data, and about 1.4 seconds for 16×16-channel protoacoustic data. Parallel computing can be used to achieve real-time denoising.

## 5. Discussion

Radiation-induced acoustic imaging is a novel modality to reveal the in-vivo radiation energy deposition. It gives rise to various techniques, such as the XACT for diagnostic imaging, the protoacoustic imaging for real-time proton range verification in proton therapy, and the EAT for electroporation monitoring. To improve their clinical utilities, it is highly desirable to reconstruct acoustic images from few-frame-averaged signals for (1) better temporal resolution with less dynamic blurriness, and (2) lower imaging dose. However, conventional averaging method or filter-based algorithms can hardly achieve satisfactory SNR from few-frame-averaged signals. In this study, we presented a deep learning-based framework to denoise few-frame-averaged acoustic signals. Results demonstrated the effectiveness and efficiency of the proposed method in achieving accurate and high-SNR signals from few-frame-averaged measurements. The acoustic image quality is considerably improved using the signals denoised by the proposed method. Both the denoised few-frame-averaged signals and the corresponding reconstructed images show very high agreement with the reference high-frame-averaged ones.

The proposed deep learning-based framework can be generalized to denoise various kinds of acoustic signals. The radiation pulse width can vary among different applications, resulting in acoustic signals of different frequencies. As a result, signal features can be presented at various temporal scales, for which inception blocks with multiple dilation rates are used in our proposed GDI-CNN. In this study, X-ray acoustic and electroacoustic signals are of mega-hertz, while protoacoustic signals are of tens of kilo-hertz due to the different beam pulse durations. As shown in the results, the proposed model achieved excellent denoising performance for all these modalities, indicating its generalizing ability for acoustic signals with large temporal characteristic variations.

The proposed method makes the radiation-induced acoustic imaging a valuable tool for both imaging and therapy monitoring. In the context of ionizing imaging, 'as low as reasonably achievable (ALARA)' principle suggests that reasonable efforts should be made to minimize the radiation exposure to patients to reduce the risk of adverse effects. As shown in results, the proposed method achieves comparable XACT image quality to 750-frame-averaged results using only 10-frame-averaged signals, considerably reducing the imaging dose by 98.7%. For the therapy monitoring, temporal resolution is an important performance metric besides accuracy. High temporal resolution not only ensures the real-time efficiency but also reduces dynamic blurriness induced by temporal averaging. Results showed that the proposed method realized proton range accuracy comparable to 1500-frame-averaged results using only 20-frame-averaged signals, which greatly improves the proton range verification frequency from 0.5Hz to 37.5Hz. And the method reached EAT image quality comparable to 750-frame-averaged results using only a single frame signal, substantially increasing the electric field monitoring frequency from 1Hz to 1kHz. Although the temporal frequency can be higher than the clinical requirement, it pushes the temporal averaging to the minimum and provides a possibility to visualize the energy deposition process with ultra-high temporal speed. These significant improvements can substantially expand the clinical utilities of these novel imaging techniques for various applications in diagnosis and treatment.

There are some limitations of this study. First, the datasets used in this study are relatively small due to the laborious data collection. Additionally, experimental data were collected from agar phantoms (containing targets) or water tanks for feasibility demonstration. In the future, more acoustic data from more complex tissues and patients are warranted to further evaluate the clinical utility of the proposed denoising method. Second, we focused on three kinds of RA signal denoising in this study due to the data availability of our labs. In the future, more imaging modalities, such as photoacoustic and MV X-ray-induced acoustic data, can be collected to further test the performance and generalizing ability of the proposed method. Third, we evaluated the results using general metrics such as element-wise errors (MSE), peak SNR (PSNR), and structural similarities (SSIM) in this study. In the future, task-specific metrics can be developed to make the evaluation more practical, which provides a more targeted assessment of the denoising performance. Furthermore, it is noted that the proposed deep learning model can experience degraded performance when further reducing the averaging frame number. Task-specific metrics can be used to explore the minimum averaging number required by the model. Fourth, it is note that, in this study, we focused on the signal denoising which is a preprocessing step for the RA imaging. The final RA image quality can be additionally improved by the advances in reconstruction and postprocessing techniques. In the future, the efficacy of the proposed denoising technique can be further evaluated in the entire chain of RA imaging.

## 6. Conclusion

The proposed deep learning-based denoising framework is a generalized method for few-frame-averaged acoustic signal denoising, which significantly improves the radiation-induced acoustic imaging's clinical utilities including low-dose XACT, real-time protoacoustic-based proton range verification, and real-time electroacoustic-based electroporation monitoring.

## Author Contributions

Dr. Lei Ren and Dr. Liangzhong Xiang are the corresponding authors who supervised and instructed the entire project.

Zhuoran Jiang led the deep learning-based method development, experiment design, and results analysis. Siqi Wang led the experiment conduction and troubleshooting of acoustic data collection. Specifically, Siqi Wang was responsible for the X-ray acoustic data collection; Yifei Xu was responsible for the electroacoustic data collection; and Gilberto Gonzalez was responsible for the protoacoustic data collection. Leshan Sun developed the codes for acoustic image reconstruction, lowpass filtering and wavelet-based denoising. Dr. Yong Chen and Dr. Q Jackie Wu provided supports for the experiment and study conduction.

All the authors contributed to the manuscript and approved the publication of this study.

## Acknowledgement

This work was supported by the National Institutes of Health under Grant No. R01-EB028324, R01-EB032680, R37CA240806, and R01-CA279013. The content is solely the responsibility of the authors and does not necessarily represent the official views of the National Institutes of Health. This work was also supported by the American Cancer Society (133697-RSG-19-110-01-CCE), and the UCI Chao Family Comprehensive Cancer Center (P30CA062203).## Data Availability

The experimental data and the source codes of the proposed denoising method will be available publicly upon acceptance of this paper.